\documentclass[twocolumn,amsmath,amssymb,superscriptaddress]{revtex4-1}

\usepackage{graphicx}
\usepackage{cleveref}
\usepackage{amssymb}
\usepackage{epstopdf}
\usepackage{bbold}
\usepackage{xcolor}

\begin{document}

\title{Electrohydraulic activity of biological cells}
\author{Marko Popovi\'c}\email{mpopovic@pks.mpg.de}
\affiliation{Max Planck Institut for the Physics of Complex Systems, N\"othnitzer Str. 38,
01187 Dresden, Germany.}
\affiliation{Center for Systems Biology Dresden, Dresden, Germany}
\author{Jacques Prost}\email{jacques.prost@curie.fr}
\affiliation{Laboratoire Physico Chimie Curie, UMR 168, Institut Curie, PSL Research University, CNRS, Sorbonne Université, 75005 Paris, France.}
\affiliation{Mechanobiology Institute, National University of Singapore, 5A Engineering Drive 1, Singapore 117411, Singapore.}
\author{Frank J\"ulicher}\email{julicher@pks.mpg.de}
\affiliation{Max Planck Institut for the Physics of Complex Systems, N\"othnitzer Str. 38,
01187 Dresden, Germany.}
\affiliation{Center for Systems Biology Dresden, Dresden, Germany}
\affiliation{Cluster of Excellence Physics of Life, TU Dresden, Dresden, Germany}

\date{\today}

\begin{abstract}
Fluid pumping and the generation of electric current by living tissues are required during morphogenetic processes and for maintainance of 
homeostasis. How these flows emerge from active and passive ion transport in cells has been well established. However, the interplay between flow and current generation is not well understood. Here, we study the electro-hydraulic coupling that arises from cell ion pumping. We develop a one-dimensional continuum model of fluid and ion transport across active cell membranes. Solving the Nernst-Planck and Poisson equations in the limit of weak charge imbalance allows us to derive approximate analytical solutions of the model. These approximations, consistent with the numerical results in physiologically relevant regime of parameters, allow us to describe electro-hydraulic activity of cells and tissues in terms of experimentally accessible parameters.
\end{abstract}

\pacs{PACS}

\maketitle

\section{Introduction}

The morphogenesis and homeostasis of tissues and organs involves the collective organisation of a large number of cells.  In general cells respond to chemical, mechanical, hydraulic, electrical and osmotic cues.
On large scales tissues are active polar materials that generate active stresses, but also fluid flows and electric currents \cite{Hodgkin1952, Piccolino2006, Zhang2006,Latorre2018, Pietak2018, Duclut2019}. This is particularly true for epithelial tissues, which have apical-basal polarity.
These behaviours can be studied in coarse-grained continuum approaches, which reveal interesting tissue properties such as fluid pumping, current generation, and electro-hydraulic coupling as well as flexo-electricity \cite{Greenebaum2018, Duclut2019} . 
In order to understand how these large scale properties of tissues result from cellular properties, it is useful to develop simple models on the scale 
of cells that capture the physics of current and flow generation.

Ion channels and pumps have been extensively studied both experimentally and theoretically \cite{Neher1976, Nelson2004, Bialek2012}.
A large body of work has been done on the generation of electric fields and electric currents in the context of action potentials and their propagation \cite{Nelson2004, Hodgkin1952}. The activity of ion pumps generate osmotic differences which give rise to water flows across membranes \cite{Ramaswamy2000, Li2015}.  Early work studied the interplay between flow and electric current generation; for example a comprehensive model of active epithelial transport was developed and applied to the transport in the Necturus gallbladder using numerical methods \cite{Weinstein1979}. 
The hydraulics of water pumping has recently raised interest in the context of the dynamics of kidney and liver but without an explicit treatment of electric currents and fields  \cite{Marbach2016, Dasgupta2018}.
In addition, the coupling of ion pumping and water flows was studied to discuss flow driven locomotion of cells through narrow channels \cite{Li2015}.
However, an analytical approach to the coupling of hydraulic and electric processes in cells is still lacking.

Here, we investigate the interplay of the electric fields and currents with fluid flows that are generated by ion pumping across membranes. We take into account the contributions of fluid flow to electric fields and ion currents in the Nernst-Planck equation and obtain a full non-linear solution close to electro-neutrality. The relative importance of fluid flow and ion diffusion is characterised by a Peclet number. We present a simplified one-dimensional setting that allows us to find exact non-linear solutions of the Nernst-Planck and Poisson equations in the limit of weak charge imbalance. 
We use this approach to first discuss a single membrane with active pumps and passive channels, both considering a closed system as well as a periodic system with a finite fluid flow and electric currents.
We relate properties of ion channels and pumps to flow velocities and electric currents and we discuss the values of these quantities obtained under physiological conditions. We show that for physiological parameter ranges, the limit of small Peclet number provides analytic expressions for transport properties which are in excellent agreement with numerical solutions of the non-linear problem.
We extend this approach to a system of two membranes which can be used as a model for a cell or an epithelium.
Eventually we show how tissue flexo-electricity can emerge from ion pumping across membranes at the cell scale.

\section{Ion densities and electric potentials in a flowing electrolyte solution}
We first discuss the effects of fluid flow in a neutral ion solution.
In thermodynamic equilibrium, the 
distances over which charges in an electrolyte interact are limited by the Debye length $\lambda_D$, due to screening by charges of opposite sign. On scales beyond the Debye length the electrolyte is essentially neutral. We consider for simplicity a one-dimensional geometry. Out of equilibrium, flows of fluid and electric currents affect  ion distributions. The current density $J^{\pm} $ of positive and negative ions
can be expressed in terms of the ion concentrations $\rho^\pm$ by the Nernst-Planck equation
\begin{align}
J^{\pm} &=  \mp K^\pm q \partial_x \phi \rho^\pm - K^\pm k_B T\partial_x\rho^\pm + 
\rho^\pm v \quad , \label{eq:NP}
\end{align}
where $K^{\pm}$ are ion mobilities, $+q$ ($-q$) is the charge of positive (negative) ions, $x$ is the distance to the membrane, $\phi$ is the electric potential and $v$ is the fluid velocity. The electric current density is $I= q(J^+ - J^-)$. In steady state, current densities are divergence free, $\partial_x J^{\pm}=0$.

The electric potential satisfies the Poisson equation $\epsilon \partial_x^2 \phi= -q(\rho^+-\rho^-)$, which together with the Nernst-Planck equation (\ref{eq:NP}) and charge conservation  
provides a one-dimensional description of the flowing electrolyte solution. 
For potential differences small compared to $k_B T/q$, the equations
can be linearised to a good approximation, and at thermodynamic equilibrium fields and charge densities
vary over the Debye length $\lambda_D$.
The solvent flow introduces an additional length scale $\lambda$ over which the ion densities vary, while maintaining charge balance. 
At large length scales the electric potential differences can exceed $k_BT/q$ in the presence of flows and nonlinearities become important (see Appendix \ref{app:Neutrality}). 

To discuss the effects of nonlinearities in the presence of flows we consider the case where charge imbalances are small,
$\vert \rho^+-\rho^-\vert \ll \rho$, where $\rho=(\rho^++\rho^-)/2$. In this case the electric potential can be approximated by imposing 
the condition $\rho^{+}= \rho^-$ in Eq. (\ref{eq:NP}) while the small charge imbalance can be calculated from the electric potential using the 
Poisson equation. 
Using this approximation we find (see Appendix \ref{app:Profiles})
\begin{align}\label{eq:rho}
\rho(x) = & \frac{1}{v}\frac{J^+/K^{+} + J^-/K^-}{1/K^+ + 1/K^-} + C e^{x/\lambda} \\    \label{eq:phi}\begin{split}
    \frac{q(\phi(x)-\phi(0))}{k_BT} = &-\left(\frac{K^+ - K^-}{K^+ + K^-}  +\frac{\frac{J^+}{K^{+}} - \frac{J^-}{K^-}}{\frac{ J^+}{K^{+}}+ \frac{J^-}{K^-}} \right)\frac{x}{\lambda}  \\
    &+ \frac{\frac{J^+}{K^{+}} - \frac{J^-}{K^-}}{\frac{J^+}{K^{+}} + \frac{J^-}{K^-}}\ln{\left(\frac{\rho(x)}{\rho(0)}\right)}    \end{split} 
    \end{align}
where, the length scale $\lambda=2k_BT K^+K^-/(v(K^+ +K^-))$ and $C$ is an integration constant which in a finite system of length $L$ reads
\begin{align}
  \label{eq:C}
C&=\frac{L/\lambda}{\exp(L/\lambda)-1}\left(\rho_0- \frac{1}{v}\frac{J^+/K^{+} + J^-/K^-}{1/K^+ + 1/K^-}\right),
\end{align}
where $\rho_0=L^{-1}\int_0^L dx\rho (x)$ is the average ion concentration and $\text{Pe} = L/\lambda$ is a Peclet number characterising the importance of drift relative to diffusion.
Note that the concentration difference across the system can be written as 
\begin{align}
\label{eq:34}
\rho(L) - \rho(0)&= \text{Pe}\left(\rho_0- \frac{1}{v}\frac{J^+/K^{+} + J^-/K^-}{1/K^+ + 1/K^-}\right) \quad ,
\end{align}
and the charge imbalance is given by
\begin{align}
\label{eq:chargeImbalance}
\rho^+ - \rho^-= \frac{1}{4\pi \lambda_B}\frac{\frac{J^+}{K^{+}} - \frac{J^-}{K^-}}{\frac{J^+}{K^{+}} + \frac{J^-}{K^-}}\left(\frac{\partial_x^2\rho(x)}{\rho(x)} - \frac{(\partial_x\rho)^2}{\rho(x)^2}  \right) \quad ,
\end{align}
where  $\lambda_B=  q^2/(\epsilon k_BT)$ is the Bjerrum length.

\section{Ion pumps and membrane permeability}
Fluid flows and electric currents can be generated by membranes which contain active pumps. In order to develop a simple model for the pumping activity  
we consider a membrane separating two fluid filled compartments, denoted $A$ and $B$. The membrane contains active pumps and passive channels for positive and negative ions. We describe the ion flux across the membrane by 
\begin{align}
\label{eq:ionPumps}
J^{\pm}&=J_0^{\pm}  + \nu^{\pm}\left(\mu^{\pm}_A - \mu^{\pm}_B \right)  
\end{align}
where $J_0$ is the flux due to  pump activity and the coefficients $\nu^{\pm}$ describe the permeability through passive ion channels. A non-zero  
$J_0$ corresponds to an asymmetric membrane in which pumps are oriented, which defines membrane polarity. For $J_0 > 0$ pumps on average transport ions from $A$ to $B$. In general,  
$J_0^{\pm}$ as well as the coefficients $\nu^{\pm}$ depend on concentrations of ions and other factors such as mechanical stress. 
This simplified expression is a good description under physiological conditions. It can be obtained from a more detailed models \cite{Huang2009}, as shown in Appendix \ref{app:ionPumps}. 

The chemical potentials governing membrane transport can be written as
\begin{align}
\label{eq:21}
\mu^{\pm}&= \mu^{\pm}_0 + k_BT \ln{\rho^{\pm}} \pm q \phi \quad ,
\end{align}
where $\mu^{\pm}_0$ are reference chemical potentials.

Fluid permeates through the membrane at a velocity which depends on the hydrostatic pressure difference $P_B - P_A$ and the osmotic pressure difference $\Pi_B - \Pi_A$ across the membrane 
\begin{align}
\label{eq:solventPermeation}
v&= \overline{\xi}\left(P_A- P_B - \Pi_A + \Pi_B \right) 
\end{align}
where $\overline{\xi}$ is a permeation coefficient.
The osmotic pressure is given by
\begin{align}
  \Pi = k_B T(\rho^+ + \rho^- + \rho_R)= k_B T(2 \rho + \rho_R) \quad ,
\end{align}
where $\rho_R$ is the number density of other solutes which do not cross the membrane barrier.

\section{Polar membrane with ion pumps and channels}

\subsection{Membrane in a sealed channel}

We first review the simple case of a closed system of length $L$ filled with solvent containing equal densities $\rho_0$ of two oppositely charged ion species. Initially, ions are homogeneously distributed such that $\rho(x) = \rho_0$. A membrane containing active pumps divides the system in two equal volumes labeled $A$ and $B$, as illustrated in Fig. \ref{fig:figSingle} a). Because the system is closed and the fluid is incompressible no fluid flow occurs, $v= 0$. Due to the activity of the pumps, 
ions flow across the membrane until a steady state is reached in which active and passive fluxes are balanced. In this steady state both volumes have different ion concentration $\rho_A$, $\rho_B$, $\rho_{R,A}$ and $\rho_{R,B}$ and there exists a pressure and an electric potential difference: 
\begin{align} 
  \rho_A &= \rho_0 \left[ 1 - \tanh{\left(\frac{J_0}{4 \nu^+ k_B T}  \right)}\right]\\
  \rho_B &= \rho_0 \left[ 1 + \tanh{\left(\frac{J_0}{4 \nu^+ k_B T}  \right)}\right]\\
  \begin{split}\label{eq:deltaPnoFlow}
  P_B - P_A &= 4\rho_0 k_B T \tanh{\left(\frac{J_0}{4\nu^+ k_B T}\right)}\\
  &+ k_BT \left( \rho_{R,B} - \rho_{R, A} \right)
  \end{split}\\
\phi_B - \phi_A &= \frac{J_0}{2\nu^+ q} \quad ,
\end{align}
where, for simplicity, we consider positive ion pumps $J_0^+= J_0, J_0^- = 0$ and the pump strength and the channel permeability are taken to be constant parameters. 

We estimate typical parameter values to  capture orders of magnitudes relevant in physiological conditions.
Typical concentration of $\rho_0= 5 \cdot 10^{25}m^{-3}$, corresponds to about $75mM$. 
Using the trans-membrane potential $|\phi_B - \phi_A|\simeq 75 mV$ we estimate $J_0/\nu^+\simeq 2.4\cdot 10^{-20} N m $. For these parameters $\rho_B/ \rho_A \simeq 10$. The order of magnitude of $4 \rho_0 k_B T$ is $8 \cdot 10^5 Pa$. Because biological membranes cannot sustain large pressures Eq. \ref{eq:deltaPnoFlow} sets the estimated value of the required concentration difference of solutes on the two sides $\rho_{R, A} -\rho_{R,B} \simeq 4 \rho_0$.

\subsection{Membrane in a periodic channel}\label{sec:periodicCapillary}

\begin{figure}[ht] 
\centering
\includegraphics[width=.5\textwidth]{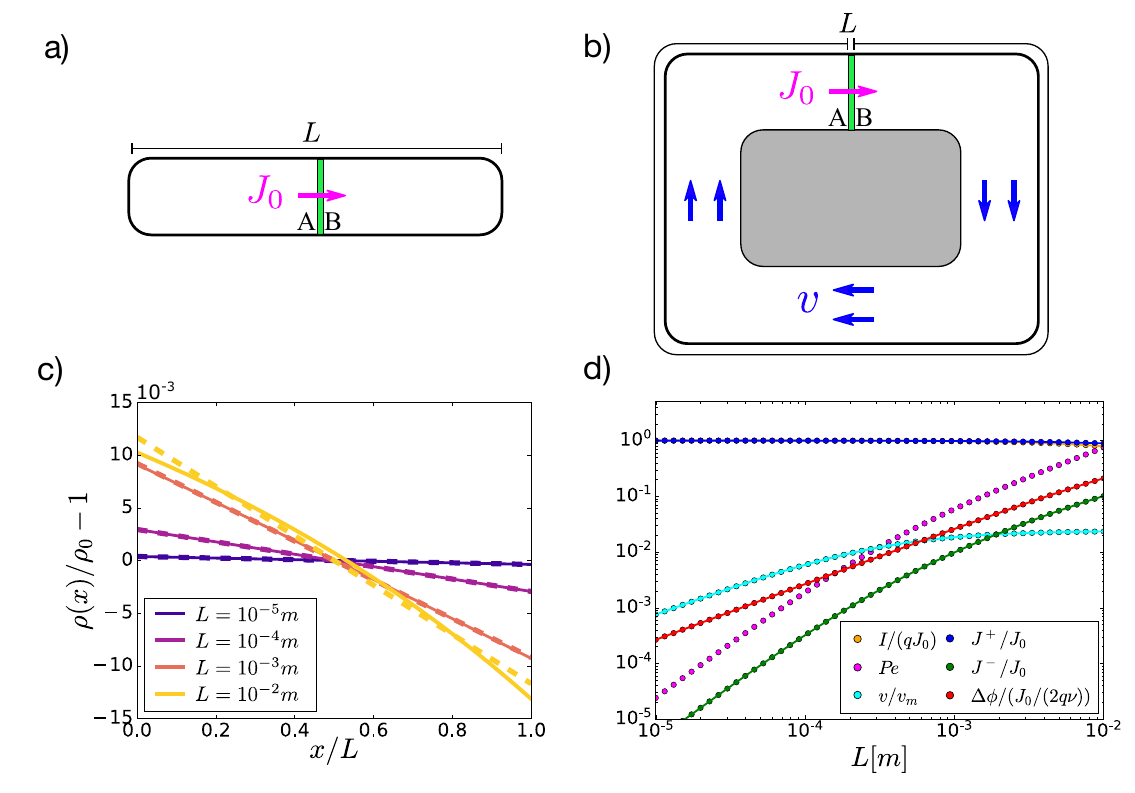}
\caption{Transport by a single active membrate. \textbf{a)} Schematic representation of an active membrane (green) with ion pumping activity $J_0$ in a sealed channel of lenght $L$. We denote the two sides of the membrane by $A$ and $B$, see text. \textbf{b)} Schematic of an active membrane in a periodic channel with the fluid velocity $v$. \textbf{c)}  Spatial profiles of density $\rho(x)$ relative to the average density $\rho_0$ as a function of relative position $x/L$ along a periodic channel. Shown are the full numerical solutions (solid lines) together with the linear approximations (see Eqs. \ref{eq:40}, \ref{eq:401}) (dashed lines).  \textbf{d)} Ion number currents $J^+$, $J^-$, electric current density $I$, electric potential across the membrane $\Delta \phi$, solvent velocity $v$, and Peclet number obtained from the full numerical solution (circles) and by our analytical approximation (lines). Ion elementary charge is denoted $q$ and $v_m$ is a maximal possible velocity, see text. Parameters used: $K= 3.1\cdot 10^{11} m^2J^{-1}s^{-1}$, $k_BT= 4.11\cdot 10^{-21}J$, $\overline{\xi}= 10^{-11}m Pa^{-1}s^{-1}$, $\rho_0= 5\cdot 10^{25}m^{-3}$, $\nu= 4.17\cdot 10^{38}kg^{-1}m^{-2}s$, $J_0=  10^{19}m^{-2}s^{-1}$.}
\label{fig:figSingle}
\end{figure}

We now consider an active membrane driving flow through a channel which forms a loop, as illustrated in Fig. \ref{fig:figSingle} b). Such a situation could be realised in a microfluidic experiment. The pumping of ions couples to fluid flow and the active membrane acts as a battery and a 
water pump. In the one-dimensional description we use periodic boundary conditions and the conditions for the current and the flow given by Eqs. \ref{eq:ionPumps} and \ref{eq:solventPermeation} at the membrane. To simplify the discussion we consider equal mobility of positive and negative ions $K= K^+= K^-$. 

For a channel filled with a gel through which a fluid permeates, the pressure changes linearly along the channel. Therefore, the pressure jump across the membrane is $P_B=P_A+v L\eta/a^2$, where $\eta$ is the fluid viscosity, $a$ is the bulk permeation length, and $v$ is the flow velocity. Using Eq. \ref{eq:solventPermeation}, the velocity can then be expressed as $v= \xi(\Pi_B - \Pi_A)$, where we introduce an effective permeation coefficient $\xi=\overline{\xi}/(1+\overline{\xi} L \eta/a^2)$.

We first derive a relation between the solvent flow and ion current densities $J^{\pm}$. From  Eq. \ref{eq:34} and Eq. \ref{eq:solventPermeation} we obtain 
\begin{align} \label{eq:flow_current}
v&= \frac{J^+ + J^-}{2\rho_0(1 + \theta)} \quad ,
\end{align}
where 
\begin{align}\label{eq:rhoStar}
\theta&= \frac{K}{2\xi L \rho_0} \quad .
\end{align}
This allows us to express ion densities at the membrane as 
\begin{align}
\label{eq:rhoA}
  \rho_A=  \rho(L)&= \rho_0 \left( 1 + \theta -  \theta\frac{e^{\text{Pe}}\text{Pe}}{e^{\text{Pe}} - 1}  \right) \quad ,\\
  \label{eq:rhoB}
\rho_B=  \rho(0)&= \rho_0 \left( 1 + \theta - \theta \frac{\text{Pe}}{e^{\text{Pe}} - 1}  \right) \quad .
\end{align}
The membrane spontaneously generates a fluid flow with velocity $v = v_m\theta \text{Pe}$, which is proportional to the Peclet number. The velocity scale $v_m = 2 \xi \rho_0 k_BT$ corresponds to the osmotic fluid flux through a membrane at concentration difference $\rho_0$. For our choice of parameters $v_m= 4- 40 \mu m/s$. Furthermore, the fluid velocity depends on the channel length via a characteristic length-scale $\theta L$. Here, we have used $K\simeq 3 \cdot 10^{11} m^2J^{-1}s^{-1}$, and range $\xi \simeq 10^{-13} - 10^{-11}m Pa^{-1} s^{-1}$, following reported values \cite{Fischbarg2010, Olbrich2000, Torres-Sanchez2021}.

In the limit of weak flow, corresponding to a small Peclet number we can find explicit solutions of the problem. For ion densities at the membrane we find
\begin{align}
\label{eq:40}
  \rho_A&\simeq \rho_0 \left( 1 - \frac{1}{2} \frac{v}{v_{\text{m}}}\right) \quad ,\\ \label{eq:401}
  \rho_B&\simeq \rho_0 \left( 1 + \frac{1}{2}\frac{v}{v_{\text{m}}} \right) \quad .
\end{align}

To determine the fluid velocity in the weak flow limit, we consider small values of $\theta\text{Pe}$, corresponding to small relative ion density variations $\ln{(\rho_A/\rho_B)} \simeq \ln{(1 - v/v_m)} \simeq -v/v_m$. For simplicity, we consider the case $\nu= \nu^{+}= \nu^-$.  Using Eq. \ref{eq:flow_current}, we then obtain
\begin{align}
\label{eq:23}
v &\simeq \frac{J_0}{2\rho_0 + \frac{\nu}{\xi\rho_0} + \frac{K}{ \xi L}} \quad .
\end{align}
Two reference concentrations are  $(\nu/2\xi)^{1/2} \simeq 45 mM$ and $K/(\xi L) \simeq 500 mM$, where we have used $\xi \simeq 10^{-11}m Pa^{-1}s^{-1}$ and $L \simeq 100 \mu m$. For values of $\rho_0$ between these reference values we can estimate the order of magnitude of the  flow velocity as  $v\simeq J_0 \xi L/K \simeq 0.3 - 30 nm /s$, where we have used $J_0\simeq 10^{17} - 10^{19}m^{-2}s^{-1}$, corresponding to an ion pump rate of $10^3 s^{-1}$ and area density of pumps of $10^2 - 10^4 \mu m^{-2}$.

We can also estimate for ion current densities and potentials considering the same limit of small $\theta \text{Pe}$. We first note that using Eq. \ref{eq:phi} the membrane potential difference $\Delta \phi_{AB} \equiv \phi_B - \phi_A$ can be expressed as 
\begin{align}
\label{eq:20}
\Delta \phi_{AB} &= \frac{k_BT}{q}\frac{J^+ - J^-}{J^+ + J^-}\left[\text{Pe} + \ln{\left( \frac{\rho_B}{\rho_A} \right)}\right] \quad .
\end{align}
Using Eqs. \ref{eq:flow_current}, \ref{eq:rhoA}, \ref{eq:rhoB} for velocity and densities we can determine from Eqs. \ref{eq:ionPumps} currents
\begin{align}
\label{eq:2}
  J^+&\simeq \frac{1}{2} \frac{1 + \theta(1 + \psi) + (1 + \theta)(1 + \psi)}{(1 +\psi)(1 + \theta(1 + \psi))}J_0 \quad , \\ \label{eq:100}
       J^-&\simeq \frac{1}{2}\frac{\psi}{(1 +\psi)(1 + \theta(1 + \psi))}J_0 \quad ,
\end{align}
where $\psi= \nu L/(K\rho_0)$.
The electric current density follows as 
\begin{align}
\label{eq:24}
I&= q\left( J^+ - J^- \right)\simeq q \frac{J_0}{1 + \psi} \quad ,
\end{align}
Finally, the membrane potential difference has the form
\begin{align}
  \label{eq:50}
  \begin{split}
 \Delta \phi_{AB}
  &\simeq \frac{J_0}{2q\nu}\frac{\psi}{1 + \psi} \quad .
  \end{split}
\end{align}

We test the accuracy of these approximations by comparing them to numerical solutions of the full non-linear problem in the weak charge imbalance limit.
Fig. \ref{fig:figSingle} c) shows numerically obtained ion density profiles along the channel, which exhibit approximately a linear profile consistent with Eqs. 19 and 20. Fig. \ref{fig:figSingle} d) shows velocity, currents and membrane potential difference as a function of the system size. It compares numerical solutions (symbols) to the analytical approximations (solid lines), showing excellent agreement. Thus, the approximations in the small $\theta\text{Pe}$ regime apply in parameter regimes consistent with physiological conditions. 

\section{Two active membranes as a simple model for cellular pumping}
\subsection{Model of cellular pumping}
Using the formalism developed above we discuss pumping across a cell by introducing two active membranes separated by the thickness $L_c$. The space between the membranes corresponds to the inside of the cell of volume $V_c$, which confines $N_R$ resident molecules.  
This could correspond for example to a cell in contact with different fluid compartments denoted $A$ and $D$, see Fig. 2 a),b). 
The resident molecules in the cell 
contribute to the cell osmotic pressure $\Pi_C= 2k_BT\rho + k_BT N_R/V_c $. 
Profile of resident molecules due to flow can be taken into account as an effective contribution to the $\xi_2$, see Appendix \ref{app:neutralProfile}, and therefore we do not account for it explicitly.   
Cell membranes exhibit an electric double layer of positive and negative charges confined within the Debye length near the membrane. In our approximation, relevant on large scales, this double layer could be taken into account as a membrane capacitance if surface charges were to be discussed. 

We first consider a symmetric cell without apical-basal polarity, see Fig. 2 a). The example shown in this figure corresponds a symmetric cell in a circular microfluidic channel. We outline how ion concentration, membrane potential and pressure difference across cell membrane appear in this case.
We then consider a polarised cell which generates fluid flows and electric currents, see Fig. 2 b).

\subsection{Symmetric cell}
In the symmetric cell we consider, ions are pumped outwards with pump strengths of magnitude $J_0$, see Fig. 2 a). Due to symmetry $\phi_A= \phi_D$ and $\phi_B= \phi_C$ and membrane potentials difference at the two membranes are the same $\Delta\phi_{AB}= \Delta \phi_{DC}$ with 
\begin{align}
\label{eq:25}
\Delta\phi_{AB}&= \frac{-J_0}{2 \nu q} \quad .
\end{align}
This potential difference sets the ion concentration $\rho^c$ inside the cell relative to the outside concentration $\rho^w$
\begin{align}
\label{eq:26}
\rho^c&= \rho^w e^{\frac{-J_0}{2\nu k_B T}} \quad ,
\end{align}
where we define the average ion concentrations inside and outside of the cell as $\rho_0^c$ and $\rho_0^w$, respectively. They satisfy $\rho_0^c L^c + \rho_0^w (L - L^c)= \rho_0 L$, where $\rho_0$ is imposed.
The pressure difference across the cell membrane $\Delta P \equiv P_B-P_A$ arises from the difference in osmotic pressures 
\begin{align}
\label{eq:33}
\Delta P& = k_BT \left[ \frac{N_R}{V_c} + 2\rho^w \left(e^{\frac{-J_0}{2 \nu k_B T}} - 1\right) \right] \quad .
\end{align}
Mechanical equilibrium requires balance of pressures which is achieved by adjusting the cell size to
\begin{align}
\label{eq:1}
L_c\simeq \frac{N_R}{2 \rho_0 A_c(1 - e^{-J_0/(2\nu k_B T)})} \quad ,
\end{align}
where $A_c$ is the area of the channel cross-section.

\subsection{Asymmetric cell in a periodic channel}
\begin{figure}[ht]
\centering 
\includegraphics[width=.45\textwidth]{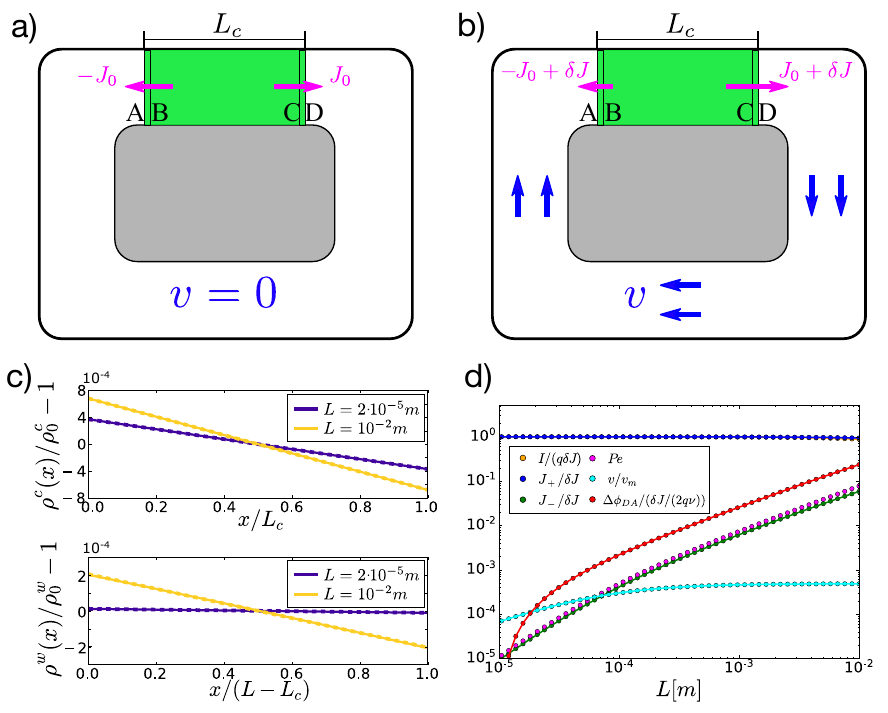}
\caption{Two membranes in a periodic channel as a model for a cell. \textbf{a)}, \textbf{b)} A schematic of a symmetric and asymmetric pair of active membranes, respectively, separated by a distance $L_c$. The membrane pumping strength is denoted $J_0$ and the asymmetry $\delta J$. The fluid velocity is denoted $v$, the sides of the two membranes are A, B, C and D.   Ion density profiles in the cell $\rho^c(x)$ (top) and outside $\rho^w(x)$ (bottom) as a funcion of normalised positions. Shown are the full numerical solutions (solid lines) together with the linear approximations Eqs. \ref{eq:42} and \ref{eq:142} (dashed lines). \textbf{d)} Ion number currents $J^+$, $J^-$, electric current density $I$ , electric potential across the cell $\Delta \phi_{DA}$, fluid velocity $v$ and the Peclet number obtained from the full numerical solutions (circles) are shown together with approximations Eqs. \ref{eq:48} - \ref{eq:51} (lines). Parameters used are the same as in Fig. \ref{fig:figSingle}, with $\delta J= 0.1 J_0$.}
\label{fig:figDouble}
\end{figure}

We now consider a polarised cell in a periodic channel of length $L$ with overall amount of ions $L \rho_0$. Inspired by the apical-basal cell polarity the outwards pumping strengths are slightly biased by an amount $\delta J$, see Fig. 2 b). Note that Eqs. \ref{eq:rho}, \ref{eq:phi}, \ref{eq:C} for density profiles and electric potential still apply, separately inside and outside of the cell. 
The membrane fluxes are described by six equations that can be solved for unknowns $J^+$,$J^-$, $v$, $L^c$, $\rho_0^c$ and $\phi_B - \phi_A$, see SI for details, from which electric potentials $\phi_{A,B, C, D}$ and ion densities $\rho_{A,B,C,D}$ follow. Spatial profiles of ion densities obtained numerically using this approach are shown in Fig. \ref{fig:figDouble} c). Currents, flow velocities and electric potentials as a function of system length $L$ are shown in Fig. \ref{fig:figDouble} d).

In order to obtain explicit expressions for these quantities we consider the limit of weak bias $\delta J \ll J_0$ where the Peclet number is small $\text{Pe} \ll 1$. This allows us to approximate the density profiles to lowest order in $\text{Pe}$ as
\begin{align}
\label{eq:42}
  \frac{\rho^c(x)}{\rho_0^c}&\simeq 1 + \text{Pe}\frac{L^c}{L} \left( 1 - \frac{\rho_0}{\rho_0^c}(1 + \theta_2) \right)\left( \frac{x}{L^c} - \frac{1}{2} \right) \quad ,\\
\label{eq:142}
    \frac{\rho^w(x)}{\rho_0^w}&\simeq 1 + \text{Pe}\frac{L - L^c}{L} \left( 1 - \frac{\rho_0}{\rho_0^w}(1 + \theta_2) \right)\left( \frac{x- L^c}{L- L^c} - \frac{1}{2} \right) 
\end{align}
where we have defined $\theta_2= K/(\rho_0 L \xi_2)$, with $2/\xi_2= 2/\bar{\xi} + (L - L^c)\eta^w /(a^w)^2 + L^c \eta^c/(a^c)^2$. Here, $a^c$ and $a^w$ are the permeation lengths in the cell and surrounding medium. 
Even inside a cell, where we estimate $a^c\simeq 5nm$  \cite{Delarue2018}, the contribution from permeation in the cell cytoplasm $L^c \eta^c/(a^c)^2 \simeq 4 \cdot 10^8 m Pa^{-1}s^{-1}$ is typically smaller than that of the cell membrane $2/\overline{\xi} \simeq 2 \cdot 10^{11}m Pa^{-1}s^{-1}$ to the effective overall permeation coefficient $\xi_2$.

The average ion density inside the cell is given by
\begin{align}
\label{eq:45}
\rho_0^c&\simeq \rho_0^w e^{\frac{-J_0}{2 \nu k_B T}}  \quad.
\end{align}
with
\begin{align}
\label{eq:47}
\rho_0^w&\simeq \frac{\rho_0}{1 - \frac{L^c}{L} \left( 1 - e^{\frac{-J_0}{2\nu k_B T}} \right)}
\end{align}
The force balanced cell size is determined by the average pump strength $J_0$
\begin{align}
\label{eq:46}
L^c&= \frac{L}{1 - e^{\frac{-J_0}{2 \nu k_B T}}}\frac{N_R}{N_I + N_R} \quad ,
\end{align}
with $A_c$ the channel cross-sectional area and $N_I= 2\rho_0 A_c L$ total number of ions in the channel. Here, as for the symmetric cell, we neglect pressure difference across the membranes as compared to osmotic pressure differences.

Solvent flow velocity and electric current density can now be determined as
\begin{align}
\label{eq:48}
  v&\simeq \frac{\delta J/(2 \rho_0)}{1 + \theta_2 +  \frac{L \nu}{2 K \rho_0} \left((1 + \theta_2)\frac{\rho_0}{L} \left( \frac{L^c}{\rho_0^c} + \frac{L - L^c}{\rho_0^w}\right) - 1 \right)} \\ 
 \label{eq:101}
  I &\simeq \frac{q\delta J}{1 + \frac{\nu}{2 K} \left( \frac{L^c}{\rho_0^c} + \frac{L - L^c}{\rho_0^w}  \right)}  \quad .
\end{align}
The membrane potentials at the two membranes 
\begin{align}
  \label{eq:49}
  \begin{split}
   \Delta \phi_{AB}&= -\frac{J_0}{2 q \nu} - \frac{J^+ - J^- - \delta J}{2 q\nu} 
  \end{split} \\
  \begin{split}
  \Delta \phi_{DC}&= -\frac{J_0}{2 q \nu} + \frac{J^+ - J^- + \delta J}{2 q\nu}
  \end{split}
\end{align}
exhibit small corrections to the value $-J_0/2q\nu$ found in the symmetric cell. 

Finally, the electric potential change across the cell is
\begin{align}
  \label{eq:51}
  \begin{split}
    \Delta \phi_{DA}    &\simeq -\frac{\frac{L - L^c}{K}\frac{\rho_0}{\rho_0^w}(1 + \theta_2)}{1 + \frac{\nu}{2 K } \left( \frac{L^c}{\rho_0^c} + \frac{L - L^c}{\rho_0^w}\right)}\frac{\delta J}{2q \rho_0 (1 + \theta_2)} \quad .
    \end{split}
\end{align}

These approximate solutions are shown in Fig. \ref{fig:figDouble} d) as solid line, which shows that they are in excellent agreement with numerical solutions of the non-linear equations.

\section{Discussion}

We have presented a one-dimensional model of electrohydraulic pumping of ions and water across membranes using the approximation of weak charge imbalance. At lengths large compared to the Debye length-scale this approximation is accurate with charge imbalances $|\rho^+ - \rho^-|/(\rho^+ + \rho^-) < 10^{-12}$ for parameter values in the physiological range that we use here.
Using this weak charge imbalance limit we present a non-linear solution of the corresponding equations in the presence of advection 
and discuss the electrohydraulic activity of a single membrane and that of a pair of membranes, capturing hydroelectric activity of a cell or of an epithelium.

Our work is related to previous work concerning the migration of cells in channels in the presence of electric fields \cite{Li2015}. Here, we discuss endogenously generated electric fields rather than a response to an externally imposed field. In addition, we provide analytical expressions for ion densities, fluxes and currents, which we show hold to a very good approximation in the physiologically relevant range. 
The geometry discussed is fundamentally different from previous work on galvanotaxy where an external electric filed is oriented parallel to the membrane \cite{Allen2013, Saw2022}.

\begin{figure}[ht] 
\centering
\includegraphics[width=.45\textwidth]{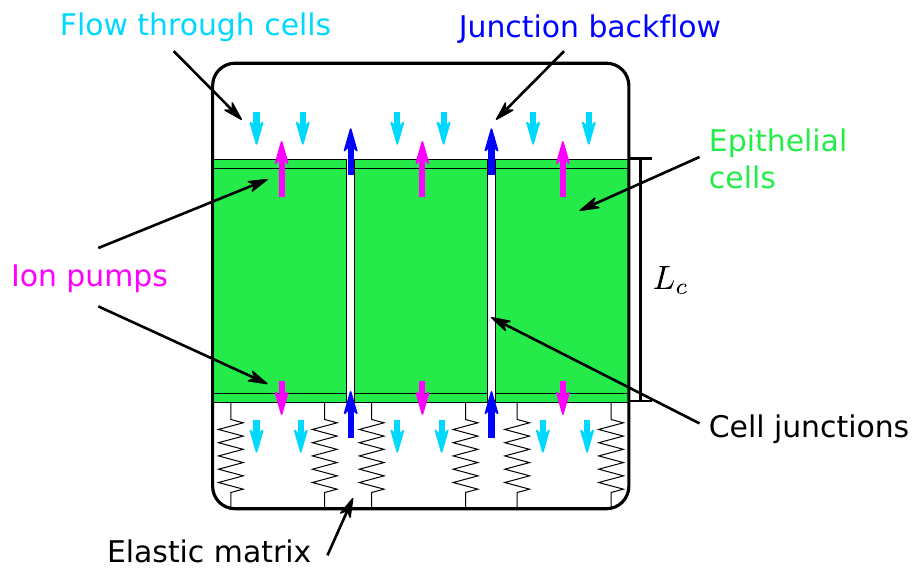}
\caption{Simple scenario of epithelial pumping. Schematic of three epithelial cells (green) separated by cell junctions. Each cell contains an asymmetric pair of membranes (magenta arrows). In the steady state the flow through the cells (light blue arrows) is balanced by the junctional backflow (dark blue arrows). Hydraulic activity leads to pressure differences balanced by the elasticity of the extracellular matrix (black springs). This scenario does not capture lateral currents and flows.}
\label{fig:figDiscussion}
\end{figure}

We have discussed a pair of membranes at a distance $L_c$ as a model for a cell in a microfluidic channel which generates an electric field and fluid flow. The same model can be interpreted as a model for an epithelium where fluid flow through the channel corresponds to the backflow of fluid through the cell-cell junctions of the epithelium where we do not account for lateral currents and flows, as shown in Fig. \ref{fig:figDiscussion}, where junctional backflow indicated by the dark blue arrows correspond to the channel flow.
This flow leads to a pressure buildup across the epithelium. 
This pressure difference drives the backflow through the cell-cell junctions and a steady state is reached when  the cell induced flow is balanced by the junction backflow.
In this case the steady state pressure difference across the epithelium is
$\Delta P\simeq \eta v d L_c/w^3\simeq 2 \cdot 10^2 Pa$, where $d \simeq 10 \mu m$ is the linear cell dimension in the epithelial plane and $w \simeq 50 nm$ is the width of cell-cell junctions, in agreement with recent experimental measurements of transepithelial pressure  \cite{Latorre2018, Choudhury2019}. This pressure difference across the epithelium can be balanced for example by a gel substrate attached to the epithelium such as an extracellular matrix layer e.g. basement membrane, or by Laplace pressure across an epithelium that bends into a dome-like shape. 

In many in-vivo cases, such as kidney \cite{Marbach2016}, the epithelium maintains a steady state flow. Also, the water balance in fresh water fish is achieved by maintaining water influx through gills and outflux through the intestinal epithelium \cite{Madsen2015}. For parameters values in the physiological range, we estimate a fluid transport velocity across an epithelium on the order of $\simeq 1 - 10  nm/s$, where we use $\delta J /J_0 \simeq 0.1$, see Eq. 36.

In addition to the hydraulic properties, the epithelium is also an electric current generator, see Eq. \ref{eq:101}. The effects of this current is analogous to those of the flow. The current first leads to a buildup of electric potential difference, corresponding to charging of a capacitor. This increasing potential difference drives a current through the cell-cell junctions. The steady state is reached when the cell induced current is balanced by the leak current. Our model corresponds to an epithelium with high resistivity of lateral membranes where lateral electric currents are negligible.
The observed trans-epithelial potentials are of the order of $\Delta \phi_{AD}\simeq 1 - 20 mV$. A naive estimate of the electric potential across the epithelium using Eq. 40 would lead to smaller values. This is because Eq. 40 considers a width of the channel equal to the width of the cell. Values in the $mV$ range can be understood as a consequence of reduced ion conductance across the cell-cell junctions by a geometric factor of about $w/d\simeq 0.05$. In our simplified model variable transport properties in the channel could be represented by changes of the effective ion mobility. Indeed, for reduced values of the ion mobility $K$ in our model, the pumping velocity is not significantly changed while the electric potential approaches $\delta J/(2 q\nu)$.  This allows us to estimate $\delta J/J_0\simeq \Delta \phi_{AD}/\Delta\phi_{AB} \simeq 0.02-0.2$.
The conductance across the cell-cell junction can be estimated based on the electric current per cell of the order of $q\delta J d^2\simeq 10^{-9} A$, where we use $\delta J \simeq 0.1 J_0$. Then, using the reported transepithelial potential difference of $10mV$ we obtain the resistance per junction on the order of $10^7 \Omega$. This can be compared to the resistance of a cell-cell junction $k_BT/(q^2\rho_0 (1 + \theta) v_m w d) \simeq 10^7\Omega$, which we estimate using linearsied Eq. \ref{eq:phi} and Eqs. \ref{eq:flow_current}, \ref{eq:40}, and \ref{eq:401} to describe a  narrow channel of width $w$. 
The relationship between hydraulic pumping and current generation by cells yields a prediction for the ratio of the electric current density and flow velocity $I/v \simeq 2 q \rho_0 (1 + \theta(1 + \psi))\simeq 6\cdot 10^8 C/m^{3}$, which could be tested experimentally.
It will be important to test the accuracy of the simple arguments  presented here in a one-dimensional framework using a more realistic two or three-dimensional geometry.

Bending of an epithelium will create a relative difference in cell areas on the two sides of the epithelium $\delta A/A \simeq C L^c$, where $C$ is the epithelium curvature. If the ion pump density is kept constant this leads to an asymmetry in pumping strength is $\delta J_b/J_0= \delta A/A$. 
The transepithelial current due to epithelium bending can be written as  $I= \Lambda_4 C$, where $\Lambda_4$ is the flexo-electric coefficient \cite{Duclut2019}. This current exists even for a symmetric cell layer if bending dynamics is slow compared to the turnover of ion pumps in the membrane. We can estimate it as $I\simeq q\delta J_b$ 
and find $I \simeq q L^c J_0 C$. Therefore, the flexo-electric coefficient is $\Lambda_4 \simeq q L^c J_0 \simeq 1.6 \cdot 10^{-5} A/m$, consistent with \cite{Duclut2019}, where we used $L^c = 10^{-5} m$ and $J_0= 10^{19}m^{-2}s^{-1}$. 

Here, we have presented a theory of fluid flows and electric currents driven by the ion pumping activity across membranes. We estimated orders of magnitude, relevant to pumping by biological cells. Our theory could be tested in microfluidic experiments. On the tissue scale such currents and flows give rise to electrohydraulic behaviours that could play an important role in tissue morphogenesis and homeostasis \cite{Cao2014, Daane2018, Duclut2019}. More generally, understanding electrohydraulics and its role in biological processes provides exciting challenges for future work.

\clearpage

\appendix

\section{Electrostatics of a flowing electrolyte solution}\label{app:Neutrality}

\subsection{Debye-H\"uckel approximation with  flow}
Here, we derive the ion density profiles in a flowing electrolyte by linearising the corresponding Nernst-Planck and Poisson equations presented in the main text. We consider a semi-infinite system bounded on one side by a membrane located at $x= 0$. Solvent flow is oriented perpendicular to the membrane in the direction of the $x$-axis and has magnitude $v$. We consider the boundary condition at positive infinity $\rho^\pm(x\to\infty) = \rho_0$.  In presence of a constant electric field $E(x\to\infty)= E_0$ at positive infinity the ion number currents are
\begin{align}
J^+&= K^+ q E_0\rho_0 + v\rho_0\\
J^-&= -K^- q E_0\rho_0 + v\rho_0 \quad.
\end{align}
The corresponding steady state electric current is $I= q^2\left(K^+ + K^-\right)\rho_0E_0$. 

We now linearise the Nernst-Planck and Poisson equations around the state at $x\to\infty$
\begin{align}\label{eq:linearized_NP}
 \pm \frac{q \rho_0 E_0}{k_BT}\left( \frac{\delta E}{E_0} +   \frac{\delta \rho^\pm}{\rho_0}\right)&= \partial_x \delta\rho^\pm - u^\pm \delta\rho^\pm \quad ,\\ \label{eq:linearisedPoisson}
\epsilon\partial_x \delta E&= q\left(\delta\rho^+ - \delta\rho^-\right) \quad,
\end{align}
where we have introduced $\delta E= E - E_0$, $\delta\rho^{\pm}= \rho^{\pm} - \rho_{0}$ and $u^\pm= v/(K^\pm k_B T)$. By taking a gradient of Eqs. \ref{eq:linearized_NP} together with Eq. \ref{eq:linearisedPoisson} we find
\begin{align}\label{eq:linearized_system}
\partial^2_x \delta\rho^+ - w^+ \partial_x \delta\rho^+ - \frac{1}{2}k_D^2\delta\rho^+ + \frac{1}{2}k_D^2\delta\rho^-&= 0 \quad ,\\
\partial^2_x \delta\rho^- - w^- \partial_x \delta\rho^- - \frac{1}{2}k_D^2\delta\rho^- + \frac{1}{2}k_D^2\delta\rho^+&= 0 \quad ,
\end{align}
where we have introduced $w^\pm=u^\pm\pm (qE_0)/(k_BT)$ and the Debye wavevector $k_D^2= (2\rho_0q^2)/(\epsilon k_BT)$. Using an exponential ansatz $\delta\rho^\pm \sim e^{k x}$ leads to the equation for the wavevector $k$
\begin{widetext}
\begin{align}
k^3 - \left(w^+ + w^-\right)k^2+ \left(w^+ w^- - k_D^2\right)k + \frac{1}{2}\left(w^+ + w^-\right)k_D^2&=0 \quad .
\end{align}
\end{widetext}
As expected, for $w^+ = w^-= 0$ the solutions of this equations are  $k_{1}^{(0)}= 0$ and $k^{(0)}_{2,3}= \mp k_D$.
In the regime of small solvent velocity and electric field, where $|w^\pm|\ll k_D $, we find solutions in the form $k_{i}= k_{i}^{(0)}+ \delta k_{i}$, assuming $\delta k_{i} \ll k_{D}$
\begin{align}
k_1&= \frac{u}{2} \quad ,\\
  k_2&= -k_D + \frac{u}{4} \quad , \\
  k_{3}&= k_{D} + \frac{u}{4} \quad .
\end{align}
Here, $u = u^+ + u^-$. For $u < 0$, corresponding to $v< 0$ i.e. solvent flow towards the membrane, the ion number densities are
\begin{align}
\begin{split}\label{eq:linearisedSolution}
\delta\rho^+ &= A^+ e^{-\left(k_D-\frac{u}{4}\right) x} + B^+ e^{\frac{u}{2} x} \\
\delta\rho^-&= -A^-\left[1 - \frac{w^+ - w^-}{k_D}\right]e^{-\left(k_D-\frac{u}{4}\right) x} + B^- e^{\frac{u}{2} x} \quad .
\end{split}
\end{align}
where Eqs. A3 require $A^+= A^-$ and $B^+= B^-$. The latter equality demonstrates that the ion density distribution associated with the velocity dependent length-scale $2/u$ is neutral, which corresponds to the weak charge imbalance limit we consider in the main text.

\subsection{Ion density and electric potential profiles}\label{app:Profiles}

We now apply the weak charge imbalance approximation
$\vert \rho^+-\rho^-\vert\ll (\rho^++\rho^-)$
to the problem of a single active periodic membrane presented in the main text. We write the Nernst-Planck equations as
\begin{align}
J^\pm &= k_B T K^{\pm} \left( \mp\frac{q}{k_B T}\rho\ \partial_x\phi - \partial_x \rho \right) + v \rho \quad .
\end{align}
Eliminating $\phi$ further yields
\begin{align}
  \partial_x \rho &= \frac{\rho}{\lambda} - \frac{J^+/K^+ + J^-/K^-}{2k_BT} \label{eq:drhodx}
\end{align}
where $\lambda=2k_BT K^+K^-/(v(K^+ +K^-))$. Solution of this equation
\begin{align}
  \label{eqApp:rho}
  \rho(x) &=  \frac{1}{v}\frac{J^+/K^+ + J^-/K^-}{1/K^+ + 1/K^-} + C e^{\frac{x}{\lambda}} \quad ,            
\end{align}
where $C$ is an integration constant.
Now, we use this result and the difference of the two equations to obtain
\begin{align}
\partial_x \phi &= \frac{1}{2q}\left[v \left(\frac{1}{K^+} - \frac{1}{K^-}\right) - \frac{J^+/K^+ - J^-/K^-}{\rho(x)}\right]
\end{align}
which is straightforward to integrate and we find
\begin{widetext}
  \begin{align}\label{eqApp:phi}
    \frac{q(\phi(x)-\phi(0))}{k_BT} = &-\left(\frac{K^+ - K^-}{K^+ + K^-}  +\frac{\frac{J^+}{K^{+}} - \frac{J^-}{K^-}}{\frac{ J^+}{K^{+}}+ \frac{J^-}{K^-}} \right)\frac{x}{\lambda} + \frac{\frac{J^+}{K^{+}} - \frac{J^-}{K^-}}{\frac{J^+}{K^{+}} + \frac{J^-}{K^-}}\ln{\left(\frac{\rho(x)}{\rho(0)}\right)}        \quad .
\end{align}
\end{widetext}

\subsection{Two membranes in a channel}
Here we outline the main equations required to solve the system of two active membranes in a periodic channel discussed in the main text.
Six equations describing the two active membranes read
\begin{align}
  \label{eq:polarJp1}
  J^+&= -J_0 + \delta J + \nu \left( k_BT \ln{\left( \frac{\rho_A}{\rho_B}\right)} -q \left(\phi_{B}- \phi_A\right)\right)\\
    \label{eq:polarJm1}
  J^-&= \nu \left( k_BT \ln{\left( \frac{\rho_A}{\rho_B}\right)} +q \left(\phi_{B}- \phi_A\right) \right)\\
  \label{eq:polarV1}
  v&= \bar{\xi} \left( P_A - P_B - 2 k_B T \left( \rho_A - \rho_B - \frac{N_R}{2V_c} \right) \right)
\end{align}

\begin{align}
    \label{eq:polarJp2}
  J^+&= J_0 +\delta J + \nu \left( k_BT \ln{\left( \frac{\rho_C}{\rho_D}\right)} -q \left(\phi_{D}- \phi_C\right) \right)\\
    \label{eq:polarJm2}
  J^-&= \nu \left( k_BT \ln{\left( \frac{\rho_C}{\rho_D}\right)} +q \left(\phi_D- \phi_C\right) \right)\\
    \label{eq:polarV2}
  v&= \bar{\xi} \left( P_C - P_D - 2 k_B T \left( \rho_C + \frac{N_R}{2V_c}- \rho_D \right) \right)
\end{align}

Now, we sum the two water flux equations to obtain
\begin{align}
\label{eq:13}
\frac{2v}{\bar{\xi}} &= P_A - P_D+ P_C - P_B -2 k_B T \left( \rho_A - \rho_B + \rho_C - \rho_D\right) \quad ,
\end{align}
where pressure differences inside and outside of the cell are $P_A - P_D= -v L^w\eta^w/(a^w)^2$ and $P_C - P_B= - vL^c \eta^c/(a^c)^2$. We expect any pressure difference across the membranes to be negligible as compared to osmotic pressure differences.  Therefore, we find
\begin{align}
\label{eq:17}
v&= - \frac{k_BT}{\xi}\left( \rho_A - \rho_B + \rho_C - \rho_D  \right) \quad ,
\end{align}
where $2/\xi_2= 2/\bar{\xi} + L^w\eta^w/(a^w)^2 +L^c \eta^c/(a^c)^2$. 
Now, adding up all ion flux equations and applying Eq. \ref{eq:17} we find
\begin{align}
\label{eq:18}
J^+ + J^-&= 2 v \rho_0 \left(1 +\theta_2\right)  \quad ,
\end{align}
where $\theta_2= K/(\xi L \rho_0)$.
Note that we do not explictly account for the osmotic contributions of the resident molecules since they can be included in the term $\xi_2$, as explained in Appendix \ref{app:neutralProfile} below.

We now rewrite the original six membrane equations as remaining five independent equations
\begin{align}
\label{eq:41}
  J^+ + J^-&= \delta J + \nu k_B T \ln{\left( \frac{\rho_A\rho_C}{\rho_B\rho_D} \right)}\\
  \label{eq:eqForCellSize}
  \rho_A + \rho_D&= \rho_B + \rho_C + \frac{N_R}{V_c}\\
  \label{eq:productTwoSides}
  \rho_B\rho_C&= \rho_A\rho_D e^{\frac{-J_0}{\nu k_B T}}\\
  J^+ - J^-&= \delta J + q \nu \left( \phi_C - \phi_B + \phi_A - \phi_D \right)\\
  J_0&= -q \nu \left( \phi_C + \phi_B - \phi_A - \phi_D \right) \quad .
\end{align}

Now, recalling that
\begin{align}
\label{eq:6}
  \frac{q (\phi_C - \phi_B)}{k_B T}&= \frac{J^+ - J^-}{J^+ + J^-}\left(\ln{\left( \frac{\rho_C}{\rho_B} \right)} -Pe^c\right)  \quad ,\\
  \frac{q (\phi_A - \phi_D)}{k_B T}&= \frac{J^+ - J^-}{J^+ + J^-}\left(\ln{\left( \frac{\rho_C}{\rho_B} \right)} -Pe^w\right)  \quad ,
\end{align}
we find an simple expression for ion number currents difference as a function of solvent velocity
\begin{align}
  \label{eq:7}
  \begin{split}
  J^+ - J^-&= \frac{(J^+ + J^-)\delta J }{\nu k_B T \text{Pe} + \delta J} \\
  &= \frac{2 \rho_0 (1 + \theta_2) v \delta J}{\frac{\nu L}{K} v + \delta J}
  \end{split}
\end{align}

To obtain analytical solution of these equations we now proceed by considering the limit $\delta J \ll J_0$ and $\text{Pe} \ll 1$ in which we expand ion density profiles to the lowest order and obtain Eqs. \ref{eq:42} and \ref{eq:142} in the main text.

\subsection{Contribution of neutral ion species to osmotic pressure differences}\label{app:neutralProfile}
Current of neutral molecules $J_R$ in a fluid flowing at velocity $v$ follows
\begin{align}
\label{eq:14}
J_R&=  -k_BT K_R \partial_x \rho_R + v \rho_R \quad ,
\end{align}
where $\rho_R$ and $K_R$ are density and mobility of the molecules, respectively.
Therefore, their spatial profile is of the form
\begin{align}
\label{eq:22}
\rho_R(x)&= \frac{J_R}{v} + C_R e^{\frac{x}{\lambda_R}} \quad ,
\end{align}
where $C_R$ is an integration constant and $\lambda_R= k_B T K_R/v$. For resident molecules in a cell which cannot leave the cell $J_R= 0$ and the ratio of $\rho_R$ at the two sides of a cell of length $L^c$ is $\exp{(\text{Pe}_R)}$, where $\text{Pe}_R= L_c/\lambda_R$ is the Peclet number of resident molecules. 

Contribution from resident molecules to osmotic pressure differences discussed below Eq. \ref{eq:13}  would be represented by a term in Eq. \ref{eq:13} that is linear in fluid velocity
\begin{align}
  \begin{split}
\label{eq:35}
  k_BT \left(\rho_R(0) - \rho_R(L^c)\right)&= - k_BT\frac{N_R L^c}{\lambda_RV}\\
  &=-\frac{v L^c N_R}{K_R V} \quad.
  \end{split}
\end{align}
Therefore, this effect can be absorbed in the effective value of $\xi_2$ as
\begin{align}
\label{eq:37}
\frac{2}{\xi_2}= \frac{2}{\overline{\xi}} + \frac{L^w \eta^w}{(a^w)^2} + \frac{L^c \eta^c}{(a^c)^2} + \frac{N_R L^c}{K_R V} \quad .
\end{align}
 Using $\rho_R \simeq 10^{26}m^{-3}$ ($150 mM$), $L^c \simeq 10^{-5} m$ and mobility of the ion species (which might be too high for large molecules) $K_R \simeq 3 \cdot 10^{11}m^2 J^{-1}s^{-1}$ we find $N_R L^c/(K_R V) \simeq 3 \cdot 10^9 m Pa^{-1}s^{-1}$, which is well below the lowest estimated value of $2/\overline{\xi} =2 \cdot 10^{11}$ we use in the main text.

\subsection{Contribution of electroosmotic effects to cellular pumping}
\label{app:electroosmosys}
Here we take into account electroosmotic effects and we solve the system of equations describing the two active membranes in a periodic channel to estimate their contribution to the overall flow.
Six equations that describe the ion and water fluxes across the two membranes read
\begin{align}\label{eq:Jp1}
  J^+&= -J_0 + \delta J + \nu \left[ k_BT \ln{\frac{\rho_A}{\rho_B}} - q \left( \phi_B - \phi_A \right) \right]\\ \label{eq:Jm1}
  J^-&=  \nu \left[ k_BT \ln{\frac{\rho_A}{\rho_B}} + q \left( \phi_B - \phi_A \right) \right]\\ \label{eq:v1}
  v&= \overline{\xi} \left[ P_A - P_B - 2 k_B T \left( \rho_A - \rho_B - \frac{N_R}{2 V_c} \right) \right]\\ \label{eq:Jp2}
  J^+&= J_0 + \delta J + \nu \left[ k_BT \ln{\frac{\rho_C}{\rho_D}} - q \left( \phi_D - \phi_C \right) \right]\\ \label{eq:Jm2}
  J^-&=  \nu \left[ k_BT \ln{\frac{\rho_C}{\rho_D}} + q \left( \phi_D - \phi_C \right) \right]\\ \label{eq:v2}
  v&= \overline{\xi} \left[ P_C - P_D - 2 k_B T \left( \rho_C - \rho_D + \frac{N_R}{2 V_c} \right) \right]  
\end{align}
Bulk solvent flow with electroosmosis are given by
\begin{align} \label{eq:bulkV1}
  v&= -(P_C - P_B)\frac{a_c^2}{\eta^c L^c}- \gamma \frac{\phi_C-\phi_B}{L^c} \quad ,\\ \label{eq:bulkV2}
  v&= -(P_A - P_D)\frac{a_w^2}{\eta^w L^w} \quad 
\end{align}
Electric potential differences are given by
\begin{align} \label{eq:elPot1}
  \frac{q}{k_B T}(\phi_C - \phi_B)&= \frac{J^+ - J^-}{J^+ + J^-} \left[ \ln{\frac{\rho_C}{\rho_B}} - \text{Pe}^c \right]\\ \label{eq:elPot2}
  \frac{q}{k_B T}(\phi_A - \phi_D)&= \frac{J^+ - J^-}{J^+ + J^-} \left[ \ln{\frac{\rho_A}{\rho_D}} - \text{Pe}^w \right]  
\end{align}
Ion density differences are given by
\begin{align} \label{eq:deltaRho1}
  \rho_C - \rho_B&= \text{Pe}^c \left( \rho_0^c - \frac{J^+ + J^-}{2 v} \right)\\ \label{eq:deltaRho2}
  \rho_A - \rho_D&= \text{Pe}^w \left( \rho_0^w - \frac{J^+ + J^-}{2 v} \right)  
  \end{align}

  Sum of ion currents can be written  in two ways, first using Eqs. \ref{eq:Jp1}, \ref{eq:Jm1}, \ref{eq:Jp2} and \ref{eq:Jm2} 
\begin{align}
\label{eq:A1}
J^+ + J^-&= \frac{1}{\alpha_1}\left(\frac{\nu L}{K} v +  \delta J \right)
\end{align}
where
\begin{align}
\label{eq:39}
  \alpha_1&= 1 + \psi^c + \psi^w\\
  \psi^{c,w}&= \frac{\nu L^{c,w}}{2 K \rho_0^{c,w}}
\end{align}
and second using Eqs. \ref{eq:v1}, \ref{eq:v2}, \ref{eq:deltaRho1} and \ref{eq:deltaRho2}
\begin{align}
\label{eq:A2}
J^+ + J^- &= 2 \rho_0 v \left( 1 + \theta_2 \right) + \frac{K \eta^c}{L a_c^2}\gamma \left( \phi_C - \phi_B \right) 
\end{align}
where
\begin{align}
\label{eq:55}
\theta_2&= \frac{K}{\xi_2 L \rho_0}
\end{align}

Now, we can write density differences as:
\begin{align}
\label{eq:A3}
  \rho_C - \rho_B&= \alpha_3 v +\beta_3 \delta J\\ \label{eq:A4}
  \rho_A - \rho_D&= \alpha_4 v + \beta_4 \delta J                   
\end{align}
where
\begin{align}
\label{eq:38}
  \alpha_3&= \frac{\rho_0^c}{T K }\left(  L^c - \frac{\psi^c L}{\alpha_1}\right)\\  
  \beta_3&= - \frac{L^c}{2 T K \alpha_1} \\
  \alpha_4&= \frac{\rho_0^w}{T K}\left(  L^w - \frac{\psi^w L}{\alpha_1}\right)\\    
  \beta_4&= - \frac{L^w}{2 T K \alpha_1}  
\end{align}

Linearising the logarithms of densities in electric potential equations we find
\begin{align}
\label{eq:A5}
J^+ - J^-&= \delta J\frac{\alpha_5 v + \beta_5 \delta J}{\overline{\alpha}_5 v + \overline{\beta}_5 \delta J}
\end{align}
where
\begin{align}
\label{eq:53}
  \alpha_5&= \frac{\nu L}{\alpha_1 K} \\
  \beta_5&= \frac{1}{\alpha_1}\\
  \overline{\alpha}_5&= \frac{\nu L}{K \alpha_1}(\alpha_1 - 1)- \nu T \left( \frac{\alpha_3}{\rho_0^{c}} + \frac{\alpha_4}{\rho_0^w} \right)= \frac{\nu L}{K}\\
  \overline{\beta}_5&= \frac{1}{\alpha_1} - \nu T \left( \frac{\beta_3}{\rho_0^c} + \frac{\beta_4}{\rho_0^w} \right) = 1 \quad .
\end{align}
We next note that this leads to a simpler expression
\begin{align}
\label{eq:58}
J^+ - J^-&= \frac{1}{\alpha_1}\delta J \quad .
\end{align}

Finally, we can use Eq. \ref{eq:elPot1} in Eq. \ref{eq:A2} and eliminate ion current sum and difference using Eqs. \ref{eq:A1}, \ref{eq:A5} as well as ion desnity difference using Eq. \ref{eq:A3} to find
\begin{align}
\label{eq:56}
v&= \frac{1 + \frac{\eta^cL^c}{2 a_c^2 \rho_0^c L}\frac{\gamma}{q}}{2 \alpha_1 \rho_0(1+\theta_2) - \frac{\nu L}{K}}\delta J \quad .
\end{align}
Here, the relative contribution from the electroosmosis can be estimated by evaluating $\eta^c L^c \gamma/(2 a_c^2 \rho_0^c L q) \approx 0.025$, where we have used $L= L_c$, $a_c= 5nm$, $\rho_0= 5 \cdot 10^{25} m^{-3}$ and $\gamma = 10^{-8}m^2/(V s)$ \cite{Saw2022,Riveline1998}. Therefore, we can neglect the contribution of electroosomosis in the calculation of the flow generated by pumping of ions across a membrane.

\section{Model for ion pumps}\label{app:ionPumps}
We introduce a simple model for ion transport across a membrane that contains active ion pumps. A single pump in this model consist of two membrane states $A$ and $B$. Ions can bind from the solution on one side of membrane to the state $A$ and from the other side of the membrane to the state $B$. Corresponding binding and unbinding rates $k^{\text{On}}_{A,B}$ and $k^{\text{Off}}_{A,B}$ satisfy detailed balance
\begin{align}
  \frac{\rho_{A,B}k^{\text{On}}_{A,B}}{k^{\text{Off}}_{A,B}}&= e^{-\frac{\Delta E_{A,B} - \mu_{A,B}}{k_BT} } \quad ,
\end{align}
where $\Delta E_{A,B}$ are energies of binding to the membrane and $\mu_{A, B}$ is a chemical potential. We denote the transition rates between states $A$ and $B$ as $r(A\rightarrow B)= r_{AB}$ and $r(B\rightarrow A)= r_{BA}$. States $A$ and $B$ can contain either one or zero ions and therefore there are four possible states of the system $(a, b)$, $a,b \in (0, 1)$. Probabilities of finding the system in one of these states are determined by the transition rates
\begin{widetext}
\begin{align}
\label{eq:200}
  \partial_t P_{0,0}&= -\left(\rho_Ak^{\text{On}}_A+ \rho_{B}k^{\text{On}}_B\right)P_{0,0} + k^{\text{Off}}_A P_{1,0} + k^{\text{Off}}_B P_{0,1}\\
  \partial_t P_{1,0}&= \rho_Ak^{\text{On}}_A P_{0,0} - \left(k^{\text{Off}}_A + r_{AB}\right)P_{1,0}  + r_{BA}P_{0,1} + k^{\text{Off}}_BP_{1,1}\\
  \partial_t P_{0, 1}&= \rho_Bk^{\text{On}}_{B}P_{0,0} + r_{AB}P_{1,0} -\left(k^{\text{Off}}_B + r_{BA}\right)P_{0,1} + k^{\text{Off}}_AP_{1,1}\\
\partial_t P_{1, 1}  &= \rho_Bk^{\text{On}}P_{1,0} + \rho_{A}k^{\text{On}}P_{0, 1} - \left(k^{\text{Off}}_{A} + k^{\text{Off}}_{B}\right)P_{1,1} \quad .
\end{align}
\end{widetext}

This system of equations has 3 degrees of freedom because 

\begin{align}
\label{eq:4}
  P_{0,0} + P_{1, 0} + P_{0, 1} + P_{1, 1}&= 1 \quad .
\end{align}

The ion transport rate can be expressed as

\begin{align}
\label{eq:5}
J_a&= P_{1,0}r_{AB} - P_{0, 1}r_{BA} \quad .
\end{align}

In the steady state we find

\begin{widetext}
  \begin{align}\label{eq:generalPumpSolution}
J_a= \frac{ \rho_A k^{\text{Off}}_B k^{\text{On}}_A r_{AB} - \rho_B k^{\text{Off}}_A k^{\text{On}}_B r_{BA}}{k^{\text{Off}}_A k^{\text{Off}}_B+k^{\text{Off}}_A r_{BA}+k^{\text{Off}}_B r_{AB} + \rho_A k^{\text{On}}_A \left(k^\text{Off}_B + r_{AB} + r_{BA}\right) + \rho_B k^{\text{On}}_B \left(k^{\text{Off}}_A+ r_{AB}+r_{BA} \right)}
\end{align}
\end{widetext}

For simplicity we consider $k^{\text{Off}}_A=k^{\text{Off}}_B$ and $k^{\text{On}}_A=k^{\text{On}}_B$ 
\begin{align}\label{eq:equalRatesCurrentGeneral}
J_a= \frac{k^{\text{On}}}{k^{\text{Off}}}\frac{\rho_A r_{AB} - \rho_B r_{BA}}{\left[1 + \left(\rho_A + \rho_B\right)\frac{k^{\text{On}}}{k^{\text{Off}}}\right]\left[1 + \frac{r_{AB} + r_{BA}}{k^{\text{Off}}}\right]} \quad .
\end{align}

Now, we include the electric potential difference across the membrane in the model by specifying transition rates $r_{AB}$ and $r_{BA}$. When pumps in the membrane are not active and the membrane is in thermodynamic equilibrium, the transition rates satisfy 
\begin{align}
\label{eq:3}
\frac{r_{AB}^{\text{eq.}}}{r_{BA}^{\text{eq.}}}&= e^{-q\frac{\phi_B-\phi_A}{k_BT}} \quad ,
\end{align}
where $\phi_{A,B}$ are electric potentials at the two sides of membrane.

To specify the absolute values of equilibrium rates we introduce a parameter $r_0$ 
\begin{align}
r_{AB}^{\text{eq.}}&= r_0 e^{-q\frac{\phi_B - \phi_A}{2k_BT}} \\
r_{BA}^{\text{eq.}}&= r_{0} e^{q\frac{\phi_B - \phi_A}{2k_BT}}
\end{align}

Consumption of fuel by the ion pumps produces an additional contribution to the transition rates
\begin{align}
r_{AB}&= r_0 e^{-q\frac{\phi_B - \phi_A}{2k_BT}} + r_a e^{-q\frac{\phi_B-\phi_A}{2k_BT} + \frac{\Delta\mu_{\text{ATP}}}{k_BT}} \\
r_{BA}&= r_0 e^{q\frac{\phi_B-\phi_A}{2k_BT}} + r_a e^{q\frac{\phi_B-\phi_A}{2k_BT} - \frac{\Delta\mu_{\text{ATP}}}{k_BT}} \quad. 
\end{align}
Here, $\Delta\mu_{\text{ATP}}$ is the energy released by consumption of an ATP molecule and $r_a$ sets the magnitude of active pumping rate.

The description of ion pumping by Eq. \ref{eq:ionPumps} corresponding to a saturated regime, where the the pumping rate is insensitive to $\Delta\mu_{\text{ATP}}$. Such regime is obtained when transport is strongly biased, $\rho_A r_{AB} \gg \rho_B r_{BA}$ and $r_{AB}\gg r_{BA}$, and transport is limited by the ion unbinding rate, $(\rho_A + \rho_B)k^{\text{On}}/k^{\text{Off}}\gg 1$ together with $(r_{AB} + r_{BA})/ k^{\text{Off}}\gg 1$. In this limit we find
\begin{align}
\label{eq:12}
J_a \simeq \frac{\rho_A}{\rho_A + \rho_B} k^{\text{Off}}\simeq \frac{1}{2} k^{\text{Off}}\quad .
\end{align}
We use this equation as an approximation for physiological conditions where $\rho_A \simeq \rho_B$ and $r_{AB} \gg r_{BA}$. The latter follows from $\phi_B - \phi_A\ll \Delta \mu_{\text{ATP}}$ using the physiological values.
The ion pump strengths in Eq. \ref{eq:ionPumps} are therefore $J_0^{\pm}= \sigma^{\pm}J_a^{\pm}$, where $\sigma^{\pm}$ are ion pump area densities and $J_a^{\pm}$ ion transport rates.

\bibliographystyle{unsrt}
\bibliography{mpopovicBib.bib}

\begin{thebibliography}{10}

\bibitem{Hodgkin1952}
Alan~L Hodgkin and Andrew~F Huxley.
\newblock A quantitative description of membrane current and its application to
  conduction and excitation in nerve.
\newblock {\em The Journal of physiology}, 117(4):500, 1952.

\bibitem{Piccolino2006}
Marco Piccolino.
\newblock Luigi galvani’s path to animal electricity.
\newblock {\em Comptes Rendus Biologies}, 329(5–6):303–318, May 2006.

\bibitem{Zhang2006}
Y.\ Zhang and A.L.\ Greer.
\newblock {{Thickness of shear bands in metallic glasses}}.
\newblock {\em Appl.\ Phys.\ Lett.}, 89(7):071907, 2006.

\bibitem{Latorre2018}
Ernest Latorre, Sohan Kale, Laura Casares, Manuel G{\' o}mez-Gonz{\' a}lez,
  Marina Uroz, L{\' e}o Valon, Roshna~V. Nair, Elena Garreta, Nuria Montserrat,
  Ar{\' a}nzazu del Campo, and et~al.
\newblock Active superelasticity in three-dimensional epithelia of controlled
  shape.
\newblock {\em Nature}, 563(7730):203–208, Nov 2018.

\bibitem{Pietak2018}
Alexis Pietak and Michael Levin.
\newblock Bioelectrical control of positional information in development and
  regeneration: A review of conceptual and computational advances.
\newblock {\em Progress in Biophysics and Molecular Biology}, 137:52–68, Sep
  2018.

\bibitem{Duclut2019}
Charlie Duclut, Niladri Sarkar, Jacques Prost, and Frank J{\" u}licher.
\newblock Fluid pumping and active flexoelectricity can promote lumen
  nucleation in cell assemblies.
\newblock {\em Proceedings of the National Academy of Sciences},
  116(39):19264–19273, Sep 2019.

\bibitem{Greenebaum2018}
Ben Greenebaum and Frank Barnes.
\newblock {\em Bioengineering and biophysical aspects of electromagnetic
  fields}.
\newblock CRC press, 2018.

\bibitem{Neher1976}
Erwin Neher and Bert Sakmann.
\newblock Single-channel currents recorded from membrane of denervated frog
  muscle fibres.
\newblock {\em Nature}, 260(5554):799–802, Apr 1976.

\bibitem{Nelson2004}
Philip Nelson.
\newblock {\em Biological physics}.
\newblock WH Freeman New York, 2004.

\bibitem{Bialek2012}
William Bialek.
\newblock {\em Biophysics: searching for principles}.
\newblock Princeton University Press, 2012.

\bibitem{Ramaswamy2000}
Sriram Ramaswamy, John Toner, and Jacques Prost.
\newblock Nonequilibrium fluctuations, traveling waves, and instabilities in
  active membranes.
\newblock {\em Physical Review Letters}, 84(15):3494–3497, Apr 2000.

\bibitem{Li2015}
Yizeng Li, Yoichiro Mori, and Sean~X. Sun.
\newblock Flow-driven cell migration under external electric fields.
\newblock {\em Physical Review Letters}, 115(26), Dec 2015.

\bibitem{Weinstein1979}
A.M. Weinstein and J.L. Stephenson.
\newblock Electrolyte transport across a simple epithelium. steady-state and
  transient analysis.
\newblock {\em Biophysical Journal}, 27(2):165–186, Aug 1979.

\bibitem{Marbach2016}
S.~Marbach and L.~Bocquet.
\newblock Active osmotic exchanger for efficient nanofiltration inspired by the
  kidney.
\newblock {\em Physical Review X}, 6(031008), 2016.

\bibitem{Dasgupta2018}
Sabyasachi Dasgupta, Kapish Gupta, Yue Zhang, Virgile Viasnoff, and Jacques
  Prost.
\newblock Physics of lumen growth.
\newblock {\em Proceedings of the National Academy of Sciences},
  115(21):E4751–E4757, May 2018.

\bibitem{Huang2009}
Feiran Huang, David Rabson, and Wei Chen.
\newblock Distribution of the na/k pumps’ turnover rates as a function of
  membrane potential, temperature, and ion concentration gradients and effect
  of fluctuations.
\newblock {\em The Journal of Physical Chemistry B}, 113(23):8096–8102, Jun
  2009.

\bibitem{Fischbarg2010}
Jorge Fischbarg.
\newblock Fluid transport across leaky epithelia: Central role of the tight
  junction and supporting role of aquaporins.
\newblock {\em Physiological Reviews}, 90(4):1271–1290, Oct 2010.

\bibitem{Olbrich2000}
K.~Olbrich, W.~Rawicz, D.~Needham, and E.~Evans.
\newblock Water permeability and mechanical strength of polyunsaturated lipid
  bilayers.
\newblock {\em Biophysical Journal}, 79(1):321–327, Jul 2000.

\bibitem{Torres-Sanchez2021}
Alejandro Torres-S{\' a}nchez, Max Kerr~Winter, and Guillaume Salbreux.
\newblock Tissue hydraulics: Physics of lumen formation and interaction.
\newblock {\em Cells \& Development}, 168:203724, Dec 2021.

\bibitem{Delarue2018}
M.~Delarue, G.P. Brittingham, S.~Pfeffer, I.V. Surovtsev, S.~Pinglay, K.J.
  Kennedy, M.~Schaffer, J.I. Gutierrez, D.~Sang, G.~Poterewicz, J.K. Chung,
  J.M. Plitzko, J.T. Groves, C.~Jacobs-Wagner, B.D. Engel, and L.J. Holt.
\newblock mtorc1 controls phase separation and the biophysical properties of
  the cytoplasm by tuning crowding.
\newblock {\em Cell}, 174(2):338--349.e20, Jul 2018.

\bibitem{Allen2013}
Greg M. Allen, Alex Mogilner, and Julie A. Theriot.
\newblock Electrophoresis of cellular membrane components creates the
  directional cue guiding keratocyte galvanotaxis.
\newblock {\em Current Biology}, 23(7):560–568, Apr 2013.

\bibitem{Saw2022}
Thuan~Beng Saw, Xumei Gao, Muchun Li, Jianan He, Anh~Phuong Le, Supatra Marsh,
  Keng-hui Lin, Alexander Ludwig, Jacques Prost, and Chwee~Teck Lim.
\newblock Transepithelial potential difference governs epithelial homeostasis
  by electromechanics.
\newblock {\em Nature Physics}, 18(9):1122–1128, Sep 2022.

\bibitem{Choudhury2019}
Mohammad~Ikbal Choudhury, Yizeng Li, Panagiotis Mistriotis, Eryn~E. Dixon, Jing
  Yang, Debonil Maity, Rebecca Walker, Morgen Benson, Leigha Martin, Fatima
  Koroma, Feng Qian, Konstantinos Konstantopoulos, Owen~M. Woodward, and
  Sean~X. Sun.
\newblock Trans-epithelial fluid pumping performance of renal epithelial cells
  and mechanics of cystic expansion.
\newblock {\em bioRxiv}, 2019.

\bibitem{Madsen2015}
Steffen~S Madsen, Morten~B Engelund, and Christopher~P Cutler.
\newblock Water transport and functional dynamics of aquaporins in
  osmoregulatory organs of fishes.
\newblock {\em The Biological Bulletin}, 229(1):70--92, 2015.

\bibitem{Cao2014}
Lin Cao, Colin~D McCaig, Roderick~H Scott, Siwei Zhao, Gillian Milne, Hans
  Clevers, Min Zhao, and Jin Pu.
\newblock Polarizing intestinal epithelial cells electrically through ror2.
\newblock {\em Journal of Cell Science}, page jcs.146357, Jan 2014.

\bibitem{Daane2018}
Jacob~M. Daane, Jennifer Lanni, Ina Rothenberg, Guiscard Seebohm, Charles~W.
  Higdon, Stephen~L. Johnson, and Matthew~P. Harris.
\newblock Bioelectric-calcineurin signaling module regulates allometric growth
  and size of the zebrafish fin.
\newblock {\em Scientific Reports}, 8(1):10391, Dec 2018.

\bibitem{Riveline1998}
Daniel Riveline, Albrecht Ott, Frank J{\" u}licher, Donald~A Winkelmann,
  Olivier Cardoso, Jean-Jacques Lacap{\` e}re, Soffia Magn{\' u}sd{\' o}ttir,
  Jean-Louis Viovy, Laurence Gorre-Talini, and Jacques Prost.
\newblock Acting on actin: the electric motility assay.
\newblock page~6.

\end{thebibliography}

\end{document}